# Dirac equation, hydrogen atom spectrum and the Lamb shift in dynamical noncommutative spaces

S. A. Alavi [*], N. Rezaei

*Department of physics, Hakim Sabzevari university, P. O. Box 397, Sabzevar, Iran*

*s.alavi@hsu.ac.ir ; alaviag@gmail.com

**Abstract**

We derive the relativistic Hamiltonian of hydrogen atom in dynamical noncommutative spaces (DNCS or τ-space). Using this Hamiltonian we calculate the energy shift of the ground state and as well the $2P_{1/2}$, $2S_{1/2}$ levels. In all cases the energy shift depend on the dynamical noncommutative parameter τ. Using the accuracy of the energy measurement we obtain an upper bound for τ. We also study the lamb shift in DNCS. Both the levels $2P_{1/2}$ and $2S_{1/2}$ receive corrections due to dynamical noncommutativity of space which is in contrast to the non-dynamical noncommutative spaces (NDNCS or θ-space) in which the level $2S_{1/2}$ receives no correction.

**Introduction**

Recently there have been notable studies on the formulation and possible experimental consequences of extensions of the usual physical theories in the noncommutative spaces. The study on noncommutative spaces is much important for understanding phenomena at short distances beyond the present test of different physical theories. Noncommutative geometry has have great impact in diverse areas of modern physics such as cosmology, gravity, high energy and quantum physics. For a review of noncommutative field theories and noncommutative quantum mechanics, see [1-3]. The noncommutative space can be realized by the coordinate operators satisfying :

$$[x_i, x_j] = i\theta_{ij} \qquad (1)$$

where $\theta_{ij}$ is an anti-symmetric tensor. The simplest situation corresponds to the case that $\theta_{ij}$ is constant, which we call non-dynamical noncommutative space or θ-space.

In general, $\theta_{ij}$ can be a function of coordinates [4,5]. Recently [6] a generalization of noncommutative space to a position dependent space is introduced which the authors called this new version of noncommutativity as dynamical noncommutative space or τ-space. The dynamical noncommtative variables in two dimensions satisfy in the following Jacobi identities :

$$[X, Y] = i\theta(1 + \tau Y^2), \qquad [X, P_x] = i\hbar(1 + \tau Y^2),$$
$$[X, P_y] = 2i\tau Y(\theta P_y + \hbar X), \quad [Y, P_y] = i\hbar(1 + \tau Y^2),$$
$$[P_y, P_x] = 0, \qquad\qquad\qquad [Y, P_x] = 0, \qquad\qquad\qquad (2)$$

It should be noted that as $\tau \to 0$, we get the $\theta$-space commutation relations :

$$[x_0, x_0] = i\theta, \qquad [x_0, p_{x0}] = i\hbar$$
$$[x_0, p_{y0}] = 0, \qquad [y_0, p_{y0}] = i\hbar$$
$$[y_0, p_{x0}] = 0, \qquad [p_{x0}, p_{y0}] = 0 \qquad (3)$$

As mentioned in [6], the X- coordinate and the momentum $P_y$ are not Hermitian, but one can find a similarity transformation i.e., a Dyson map $\eta^{-1}O\eta = 0 = 0^\dagger$ (with $\eta = (1 + \tau Y^2)^{-1/2}$) and convert the non-Hermitian variables into a Hermitian one. The new Hermitian variables x, y, $p_x$ and $p_y$ can be expressed in terms of noncommutative $\theta$-space as follows [6] :

$$x = \eta X \eta^{-1} = (1 + \tau y_0^2)^{-\frac{1}{2}}(1 + \tau y_0^2)x_0(1 + \tau y_0^2)^{\frac{1}{2}} = (1 + \tau y_0^2)^{\frac{1}{2}}x_0(1 + \tau y_0^2)^{\frac{1}{2}}$$
$$y = \eta Y \eta^{-1} = (1 + \tau y_0^2)^{-\frac{1}{2}}y_0(1 + \tau y_0^2)^{\frac{1}{2}}$$
$$p_x = \eta P_x \eta^{-1} = (1 + \tau y_0^2)^{-\frac{1}{2}}p_{x0}(1 + \tau y_0^2)^{\frac{1}{2}} = p_{x0}$$
$$p_y = \eta P_y \eta^{-1} = (1 + \tau y_0^2)^{-\frac{1}{2}}(1 + \tau y_0^2)p_{y0}(1 + \tau y_0^2)^{\frac{1}{2}} = (1 + \tau y_0^2)^{\frac{1}{2}}p_{y0}(1 + \tau y_0^2)^{\frac{1}{2}} \qquad (4)$$

These new Hermitian variables satisfy in the same commutation relations as (2) :

$$[x, y] = i\theta(1 + \tau y^2), \qquad [x, p_x] = i\hbar(1 + \tau y^2)$$
$$[p_x, p_y] = 0, \qquad\qquad [x, p_y] = 2i\tau y(\theta p_y + \hbar x)$$
$$[y, p_x] = 0, \qquad\qquad [y, p_y] = i\hbar(1 + \tau y^2) \qquad (5)$$

Using Bopp shift one can express the noncommutative $\theta$-variables in terms of the standard (commutative) variables [7] :

$$x_{i0} = x_{i_s} - \frac{\theta_{ij}}{2\hbar}p_{j_s}, \quad p_{i0} = p_{i_s}, \quad i,j = x,y \qquad (6)$$

where $\theta_{ij} = \epsilon_{ijk}\theta_k$, one can take $\theta_3 = \theta$ and the rest of the θ-components to zero (which can be done by a rotation or a redefinition of coordinates).
It is worth mentioning that the subscript "0" indicates quantities in noncommutative θ-space, while subscript "$s$" denotes one in standard or commutative space.
The interesting point is that in the DNCS there is a minimum length for X in a simultaneous X,Y measurement [6] :

$$\Delta X_{min} = \theta\sqrt{\tau}\sqrt{1+\tau\langle Y\rangle_\rho^2} \qquad (7)$$

but there is no nonvanishing minimal length for Y. This means that objects in DNCS are naturally of string type.

**Dirac equation in the dynamical noncommutative space**

Electronic bound states around charged impurities in two dimensional systems can be described in terms of a two dimensional hydrogen atom. In this section we study the relativistic hydrogen atom in a two dimensional dynamical noncommutative space. The general form of the Dirac equation for the hydrogen atom is :

$$H = \vec{\alpha}\cdot\vec{p} + m\beta + eA_0 = \alpha_1 P_1 + \alpha_2 P_2 + m\beta + eA_0 \qquad (8)$$

where $A_0$ is given by $-\dfrac{e}{(X^2+Y^2)^{\frac{1}{2}}}$.

Using relations (4), we rewrite the Hamiltonian in terms of θ-space variables :

$$\alpha_1 p_{x0} + \alpha_2((1+\tau y_0^2)^{\frac{1}{2}}p_{y0}(1+\tau y_0^2)^{\frac{1}{2}})) + m\beta - e^2\left((1+\tau y_0^2)^{\frac{1}{2}}x_0(1+\tau y_0^2)x_0(1+\tau y_0^2)^{\frac{1}{2}} + y_0^2\right)^{-\frac{1}{2}} \qquad (10)$$

Since τ and θ are small, so the parentheses can be expanded to the first order using $(1+\tau y_0^2)^{\frac{1}{2}} = 1 + \frac{1}{2}\tau y_0^2$, then we have :

$$\alpha_1 p_{x0} + \alpha_2\left(p_{y0} + \frac{1}{2}\tau p_{y0}y_0^2 + \frac{1}{2}\tau y_0^2 p_{y0}\right) + m\beta - e^2\left(\left(x_0^2 + \tau x_0 y_0^2 x_0 + \frac{1}{2}\tau x_0^2 y_0^2 + \frac{1}{2}\tau y_0^2 x_0^2\right) + y_0^2\right)^{-\frac{1}{2}} \qquad (11)$$

Using Bobb-shift (6), we express the θ-variables $x_0, y_0, p_{x0}$ and $p_{y0}$ in terms of the standard (commutative) space variables, then Eq.(11) reads :

$$\alpha_1 p_{xs} + \alpha_2 \left( p_{ys} + \frac{1}{2}\tau p_{ys}\left(y_s - \frac{\theta_{21}}{2\hbar}p_{x_s}\right)^2 + \frac{1}{2}\tau \left(y_s - \frac{\theta_{21}}{2\hbar}p_{x_s}\right)^2 p_{ys} \right) + m\beta - e^2 \Big( \Big( x_s -$$
$$\frac{\theta_{12}}{2\hbar}p_{y_s} \Big)^2 + \frac{1}{2}\tau\left(x_s - \frac{\theta_{12}}{2\hbar}p_{y_s}\right)^2 \left(y_s - \frac{\theta_{21}}{2\hbar}p_{x_s}\right)^2 + \tau\left(x_s - \frac{\theta_{12}}{2\hbar}p_{y_s}\right)\left(y_s - \frac{\theta_{21}}{2\hbar}p_{x_s}\right)^2 \Big(x_s -$$
$$\frac{\theta_{12}}{2\hbar}p_{y_s} \Big) + \frac{1}{2}\tau\left(y_s - \frac{\theta_{21}}{2\hbar}p_{x_s}\right)^2 \left(x_s - \frac{\theta_{12}}{2\hbar}p_{y_s}\right)^2 + \left(y_s - \frac{\theta_{21}}{2\hbar}p_{x_s}\right)^2 \Big)^{-\frac{1}{2}} \qquad (12)$$

So to the first order in θ, Eq.(10) becomes :

$$\alpha_1 p_{xs} + \alpha_2 \left( p_{ys} + \frac{1}{2}\tau p_{ys} y_s^2 + \frac{1}{2}\tau y_s^2 p_{ys} \right) + m\beta$$
$$- e^2 \left( x_s^2 - \frac{\theta}{\hbar}x_s p_{ys} + \frac{1}{2}\tau x_s^2 y_s^2 + \tau x_s y_s^2 x_s + \frac{1}{2}\tau y_s^2 x_s^2 + y_s^2 + \frac{\theta}{\hbar}y_s p_{xs} \right)^{-\frac{1}{2}}$$

Employing the fundamental relation $[x_i, p_j] = i\hbar\delta_{ij}$, we rearrange the term $\frac{1}{2}\tau p_{ys} y_s^2$ as :

$$\frac{1}{2}\tau p_{ys} y_s^2 = \frac{1}{2}\tau y_s^2 p_{ys} - i\hbar\tau y_s$$

Therefore the Hamiltonian takes the following form :

$$H = \alpha_1 p_{xs} + \alpha_2 p_{ys} + m\beta + \alpha_2 \tau y_s^2 p_{ys} - \alpha_2 \, i\hbar\tau y_s - \frac{e^2}{(x_s^2 + y_s^2)^{\frac{1}{2}}} + \frac{e^2}{2(x_s^2 + y_s^2)^{\frac{3}{2}}}$$
$$\times \left( 2\tau x_s^2 y_s^2 - \frac{\theta}{\hbar}L_z \right)$$

which can be written as :

$$H = H^s + H^\tau + H^\theta$$

where :

$$H^\tau = \alpha_2 \tau y_s^2 p_{ys} - \alpha_2 \, i\hbar\tau y_s + \frac{e^2}{2(x_s^2+y_s^2)^{\frac{3}{2}}} \times (2\tau x_s^2 y_s^2), \quad H^\theta = -\frac{e^2}{2(x_s^2+y_s^2)^{\frac{3}{2}}} \times \frac{\theta}{\hbar}L_z \qquad (13)$$

are the perturbation Hamiltonians and reflect the effects of dynamical and non-dynamical noncommutativity of space on the relativistic Hamiltonian of the hydrogen atom. The effects of $H^\theta$ on Dirac equation is studied in [8], in this paper, we treat $H^\tau$, as a perturbative Hamiltonian and study its effect on Dirac equation. The energy shift of hydrogen atom due to $H^\tau$, described by Dirac equation is given by :

$$\Delta E = \int_0^\infty \int_0^\pi \int_0^{2\pi} \psi_{jm}^\dagger \, H^\tau \, \psi_{jm} \, r^2 dr \, \sin\theta \, d\theta \, d\varphi$$

The unperturbed wave functions are [9,10] :

$$\psi_{jm} = r^{\frac{-n-1}{2}} \begin{pmatrix} ig(r)\sqrt{\frac{j+m}{2j}} Y_{l,m-\frac{1}{2}} \\ ig(r)\sqrt{\frac{j-m}{2j}} Y_{l,m+\frac{1}{2}} \\ f(r)\sqrt{\frac{j+m}{2j}} Y_{l,m-\frac{1}{2}} \\ -f(r)\sqrt{\frac{j-m}{2j}} Y_{l,m+\frac{1}{2}} \end{pmatrix} \quad (15)$$

where $f(r)$ and $g(r)$ are given by :

$$\begin{Bmatrix} g(r) \\ f(r) \end{Bmatrix} = \frac{\pm (2\lambda)^{3/2}}{\Gamma(2\gamma_D + 1)} \times \sqrt{\frac{(mc^2 \pm E)\Gamma(2\gamma_D + n' + 1)}{4mc^2 \frac{(n' + \gamma_D)mc^2}{E} \left[\frac{(n' + \gamma_D)mc^2}{E} - k_D\right] n'!}}$$

$$\times (2\lambda r)^{\gamma_D} e^{-\lambda r} \left\{ \left[\frac{(n'+\gamma_D)mc^2}{E} - k_D\right] \times F(-n', 2\gamma_D + 1, 2\lambda r) \mp n' F(1 - n', 2\gamma_D + 1, 2\lambda r) \right\} \quad (16)$$

and, $F(a, c; x)$ is Hypergeometric function :

$$F(a, c; x) = 1 + \frac{a}{c}x + \frac{a(a+1)x^2}{c(c+1)2!} + \cdots$$

D is the dimension of space and :
$$\gamma_D = \left(j + \frac{D-2}{2}\right)^2 - (Z\alpha)^2$$

$$\lambda = \frac{(m^2c^4 - E^2)^{1/2}}{\hbar c}, \; j = l \pm \frac{1}{2}, \; n' = n - j - \frac{D-3}{2}.$$

Using the first order perturbation theory we calculate the shift of energy for the ground state , the result is as follows :

$$\Delta E_{ground} = (260.401 \text{ eV } m^2) \, \tau \quad (17)$$

The accuracy of the energy measurement is $10^{-12}$ eV [11], so we can put the following upper bound on the dynamical noncommutative parameter τ:

$$\tau < \frac{10^{-12}\ ev}{260.401\ eV\ m^2} = \frac{10^{-12}}{260.401\ m^2} \tag{18}$$

Using the relation $1\ Fermi \approx 5\ (GeV)^{-1}$, one can get:

$$\sqrt{\tau} \leq 10^{-17} eV \tag{19}$$

which is consistent with the one obtained in [12].
It is worth mentioning that upper and lower bounds for the non-dynamical non-commutative parameter $\theta$ were obtained in [13] and [14] respectively.

**The Lamb shift**

According to the Dirac equation in commutative space, the energy states $2S_{1/2}$ and $2P_{1/2}$ are degenerate i.e., have the same energy. $\theta$-noncommutativity has no effect on $2S_{1/2}$ level [8] and just causes energy correction to $2P_{1/2}$ level. But τ -noncommutativity affects both $2P_{1/2}$ and $2S_{1/2}$ levels and make some corrections to their energies, the situation is illustrated in Fig. (1). The energy shift for the levels $2S_{1/2}$ and $2P_{1/2}$, respectively are as follows:

$$\Delta E_{2S_{\frac{1}{2}}} = (4.87865\ eV\ m^2)\tau$$

$$\Delta E_{2P_{\frac{1}{2}}} = (7.31797\ eV\ m^2)\tau \tag{20}$$

In Ref. [8] it is shown that the $\theta$-noncommutativity in addition to the normal Lamb shift $2P_{1/2}^{1/2} \to 2S_{1/2}$ has a new channel $2P_{1/2}^{-1/2} \to 2S_{1/2}$ as well, but it has no effect on $2S_{1/2}$ level. But, in a τ-space in addition to those two channels, the energy correction of $2S_{1/2}$ level, causes further enhancement in the transition width (rate). So, high accuracy energy measurement of the $2S_{1/2}$ level and the Lamb shift may be a good criteria to determine whether noncommutativity of space is dynamical or non-dynamical.
In Ref.[15], it is shown that the energy correction of the ground state of a harmonic oscillator is also a good criteria for checking whether the noncommutativity of space is dynamical or non-dynamical. It may be also useful to mention that the noncommutative harmonic oscillator (in noncommutative $\theta$-space) has been studied in [16,17].

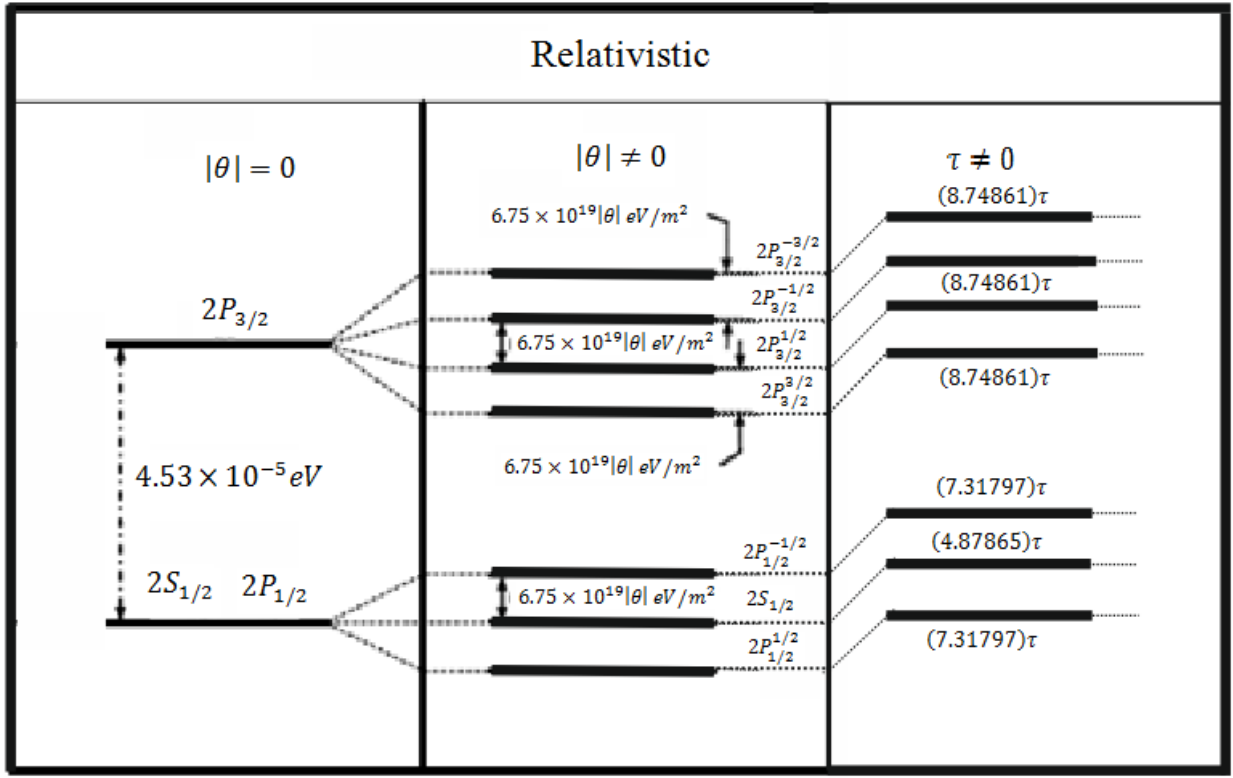

Figure 1 – Splittings for relativistic energy levels due to non-dynamical and dynamical space noncommutativity. All numerical values of the coefficients of the first-order $\tau$-corrections are in units of $eVm^2$.

**Concluding remarks**

String theory provides the first way of putting quantum theory and general relativity together and is therefore a candidate for a theory of everything. It is shown that in contrast to the non-dynamical noncommutative space the objects in the dynamical noncommutative space studied here, are string like, so it seems that there is a more strong relation between DNCS with string theory than NDNCS. On the other hand in recent years there has been a growing interest in theoretical and experimental studies of non-Hermitian operators in physics and mathematics [18]. As mentioned in the "Introduction", some operators in DNCS are non-Hermitian. This interface of DNCS and the theory of non-Hermitian operators in one hand and DNCS and string theory on the other hand can lead to fundamental new insights in all three fields. So it is interesting to study fundamental phenomena in DNCS. In this paper the Hamiltonian of the hydrogen atom described by Dirac equation in DNCS is derived and as a result, the energy corrections to the ground state as well the $2S_{1/2}$, $2P_{1/2}$ states have been all calculated. Using the energy measurment accuracy, an upper bound for the dynamical noncommutativity parameter $\tau$ has been obtained. Then it is shown that in contrast to the NDNCS, which has no effect on the $2S_{1/2}$ level, the DNCS affects both $2S_{1/2}$ and $2P_{1/2}$ levels, and subsequently causes energy shift and enhances the width of the transition. So in future high-precision spectroscopy data in combination with spectral retrieval techniques enable us to determine whether there exists any noncommutativity of space in nature and whether it is dynamical or non-dynamical.